\documentclass[sigconf]{acmart}
\AtBeginDocument{%
  }

\usepackage{algorithm,algpseudocode,placeins,array,makecell,multirow}

\begin{document}

\title{TaoSR1: The Thinking Model for E-commerce Relevance Search}


\author{Chenhe Dong}
\authornote{Equal contribution.}
\email{dongchenhe.dch@alibaba-inc.com}
\affiliation{%
  \institution{Taobao \& Tmall Group of Alibaba}
  \city{Hangzhou}
  \country{China}
}

\author{Shaowei Yao}
\authornotemark[1]
\authornote{Corresponding author.}
\email{yaoshaowei@taobao.com}
\affiliation{%
  \institution{Taobao \& Tmall Group of Alibaba}
  \city{Hangzhou}
  \country{China}
}

\author{Pengkun Jiao}
\email{pkjiao23@m.fudan.edu.cn}
\affiliation{%
  \institution{Fudan University}
  \city{Shanghai}
  \country{China}
}

\author{Jianhui Yang}
\email{yangjh23@mails.tsinghua.edu.cn	}
\affiliation{%
 \institution{Tsinghua University}
 \city{Beijing}
 \country{China}}

\author{Yiming Jin}
\email{enxian@alibaba-inc.com}
\affiliation{%
  \institution{Taobao \& Tmall Group of Alibaba}
  \city{Hangzhou}
  \country{China}}

\author{Zerui Huang}
\email{huangzerui.hzr@taobao.com}
\author{Xiaojiang Zhou}
\email{zxjbupt@163.com}
\affiliation{%
  \institution{Taobao \& Tmall Group of Alibaba}
  \city{Hangzhou}
  \country{China}}

\author{Dan Ou}
\email{oudan.od@taobao.com}
\affiliation{%
  \institution{Taobao \& Tmall Group of Alibaba}
  \city{Hangzhou}
  \country{China}}

\author{Haihong Tang}
\email{piaoxue@taobao.com}
\affiliation{%
  \institution{Taobao \& Tmall Group of Alibaba}
  \city{Hangzhou}
  \country{China}}

\author{Bo Zheng}
\email{bozheng@alibaba-inc.com}
\affiliation{%
  \institution{Taobao \& Tmall Group of Alibaba}
  \city{Beijing}
  \country{China}}

\renewcommand{\shortauthors}{Dong et al.}

\begin{abstract}
Query-product relevance prediction serves as a foundational technology in e-commerce search engines, enabling users to discover desired products and ensuring optimal user experience. Previous approaches have primarily relied on BERT-based models which excel at textual and basic semantic matching but demonstrate poor performance in understanding and reasoning capabilities for more complex queries. Consequently, numerous recent studies have explored the application of Large Language Models (LLMs) in search systems. However, most still adopt discriminative paradigms or ultimately distill knowledge to BERT models for deployment.
In this paper, we propose an optimization framework based on large language models and directly deploy these models in online systems. Nevertheless, practical deployment presents several challenges, including online deployment, error accumulation in Chain-of-Thought (CoT) leading to performance degradation, and discriminative hallucination. To address these challenges, we propose an LLM-based optimization framework called \textbf{Tao}bao \textbf{S}earch \textbf{R}elevance Model v\textbf{1} (TaoSR1) comprising three stages: (1) Supervised Fine-Tuning (SFT) with CoT to endow models with reasoning capabilities; (2) Offline multiple sampling based on a pass@N strategy, combined with Direct Preference Optimization (DPO), to enhance model generation quality; and (3) Difficulty-based dynamic sampling integrated with Group Relative Policy Optimization (GRPO) to further mitigate model's discriminative hallucination problems. Finally, by incorporating post-CoT processing and a relevance tier partitioning method based on cumulative probability, our model achieves more feasible and efficient online deployment. Experimental results demonstrate that our proposed model significantly outperforms baseline methods on challenging offline evaluation datasets, while achieving substantial improvements in online side-by-side human evaluations. Our proposed framework introduces a novel optimization paradigm for incorporating CoT reasoning into relevance classification tasks; we contend that this methodology  provide valuable insights into the application of LLMs for other classification tasks.
\end{abstract}




\maketitle

\section{Introduction}
Large-scale e-commerce platforms such as Taobao and Amazon serve hundreds of millions of users daily, tasked with retrieving desired products from a massive item corpus. Query-item relevance prediction ensures the relevance between returned products and search queries.  It is a critical determinant of user experience and a significant factor in long-term user value. Displaying irrelevant products not only negatively impacts user experience but also reduces merchant operational efficiency and satisfaction. Therefore, query-item relevance prediction technology constitutes the cornerstone of e-commerce search engines.

Previous approaches primarily relied on BERT-based models, leveraging an encoder-only architecture and bidirectional attention mechanisms to achieve strong text matching capabilities, typically satisfying 80-90\% of user search needs. However, the remaining over 10\% of long-tail complex queries present a significant challenge, often resulting in inconsistent user experiences. These long-tail queries demand stronger semantic understanding and reasoning capabilities from models. The low-rank problem inherent in BERT's bidirectional attention limits parameter scaling, thereby constraining the upper bound of BERT model capabilities. Recent research has increasingly explored the application of Large Language Model (LLM) technologies in search systems, including works such as RankLLaMA\cite{10.1145/3626772.3657951}, LREF\cite{10.1145/3701716.3715246} and so on\cite{chen-etal-2025-towards-boosting,mehrdad2024large}. However, most approaches still employ discriminative paradigms or distill knowledge to BERT models for deployment. As a result, they fall short of effectively addressing queries that require complex reasoning.

In this paper, we propose a novel framework to optimize and directly deploy a generative LLM for online relevance prediction. However, practical implementation presents significant challenges across three dimensions: (1) Deployment Latency: While Chain-of-Thought (CoT) reasoning enhances model reasoning capabilities and significantly improves the upper bound for solving complex problems, the proportional increase in output tokens correspondingly increases response latency. For relevance models, a single request requires computing relevance between a query and hundreds of candidate items, making real-time online generation computationally prohibitive. (2) Error Accumulation in CoT: Since CoT generates longer intermediate processes, a single hallucination or reasoning error at any intermediate step can propagate and affect final results, leading to incorrect classification outcomes. (3) Discriminative Hallucination: Even with a correct reasoning chain, the model may still produce an incorrect final answer.

To address these challenges, we propose an LLM-based optimization framework comprising three stages: (1) Supervised Fine-Tuning (SFT) with CoT to endow models with reasoning capabilities, improving the upper bound of model discrimination ability; (2) Offline pass@N sampling with Direct Preference Optimization (DPO) to enhance correct answering capability, reducing result errors caused by process errors; (3) Difficulty-based Dynamic Sampling with Group Relative Policy Optimization (GRPO) to further alleviate discriminative hallucination. Experimental results demonstrate that our thinking model significantly outperforms baseline models on a curated dataset of challenging queries, exhibiting stronger reasoning and comprehension capabilities. Furthermore, it achieves substantial improvements in online side-by-side human evaluations.

\section{Related Works}
\subsection{Relevance Search}
Relevance modeling is one of the most fundamental tasks in information retrieval, aiming to model the semantic relevance between queries and items. The field of relevance modeling has experienced extensive historical development. Early approaches focused on feature-based modeling, employing algorithms such as BM25\cite{svore2009machine} and TF-IDF\cite{aizawa2003information}, which relied on manually defined features to address relevance modeling challenges. While these methods proved effective in early information retrieval systems, they lacked cross-scenario generalizability and demanded substantial manual effort.
Subsequently, deep learning-based representation modeling methods\cite{shen2014learning,10.1145/3442381.3450129} emerged, which utilized deep learning models to encode queries and items as embeddings and further modeled their relevance relationships. However, these approaches lacked the capacity to model broad knowledge and thus possessed limited potential for solving complex cases, with performance remaining suboptimal.
The introduction of BERT\cite{devlin2019bert} marked a paradigm shift, making fine-tuning of pre-trained language models with Transformer\cite{vaswani2017attention} architecture a mainstream approach for relevance modeling\cite{bo2021modeling,10.1145/3534678.3539090,10.1145/3637528.3671559}. These methods demonstrated the capability to model complex semantic relationships, while their unified task paradigm and end-to-end training architecture significantly reduced the manual effort required for feature engineering.
In recent years, decoder-only large language models have exhibited exceptional natural language understanding and generation capabilities, virtually unifying upstream and downstream task paradigms across natural language processing, leading to the emergence of relevance modeling research based on large language models. Notable works including LREF\cite{tang2025lref}, ProRBP\cite{chen2024towards}, and research by Mehrdad et al.\cite{mehrdad2024large} have employed training methodologies such as supervised fine-tuning (SFT) and Direct Preference Optimization (DPO)\cite{rafailov2023direct} to endow large language models with query-item relevance modeling capabilities.
However, by adhering to discriminative optimization paradigms or distilling knowledge to smaller BERT-like models for deployment, most prior work fails to effectively address queries that require complex reasoning. In this paper, we address this gap by proposing a framework to optimize a large language model for relevance and deploy it directly in a production online setting.

\subsection{Reasoning Large Language Models}
While large language models have demonstrated exceptional capabilities across numerous downstream tasks, conventional large language models still struggle to excel in domains requiring strong reasoning abilities, such as mathematics and programming. To address this limitation, early approaches employed prompt engineering\cite{kojima2022large} and supervised fine-tuning\cite{snell2024scaling} to guide models in generating step-by-step reasoning chains, thereby improving their reasoning capabilities. However, these methods lacked cross-domain generalizability and failed to fully leverage the rich world knowledge embedded in large language models.
The emergence of o1\cite{jaech2024openai} and DeepSeek-R1\cite{guo2025deepseek} has sparked a surge in using reinforcement learning to enhance the reasoning capabilities of large language models. These reasoning models, including o1 and DeepSeek-R1, employ reinforcement learning algorithms such as PPO\cite{schulman2017proximal} or GRPO\cite{shao2024deepseekmath} to reward answers sampled from the models themselves, thereby activating the models' inherent rich world knowledge and improving their performance on challenging reasoning problems in mathematics and programming. Subsequently, additional reinforcement learning algorithms\cite{yu2025dapo}\cite{zheng2025group} and reasoning models\cite{team2025kimi}\cite{seed2025seed1} have been introduced, establishing the use of reinforcement learning algorithms on large language models to enhance reasoning capabilities as a widely adopted technical paradigm.
However, current research predominantly focuses on the two domains of mathematics and programming, with limited exploration of reinforcement learning applications in vertical industries and diverse task paradigms. Our work provides an exploratory investigation into the application of reinforcement learning in the e-commerce search domain.

\subsection{Reinforce Learning for Classification}
E-commerce search relevance tasks can be formulated as binary or multi-classification problems, where the input consists of query and item features, and the output is the model-predicted relevance label. While numerous studies have explored reinforcement learning for training large language models, most current research focuses primarily on complex generation tasks in mathematics and programming domains, with relatively limited exploration of reinforcement learning applications in generative classification tasks.
A particularly relevant work is GenCLS++\cite{he2025gencls++}, which systematically investigates the performance variations of large language models on generative text classification tasks under different generation and training paradigms. Interestingly, unlike their performance in mathematics and programming domains, in generative classification tasks, models often achieve better performance when they either do not output reasoning paths or place the reasoning process after the output label. Furthermore, conducting reinforcement learning after supervised fine-tuning can yield additional performance improvements.
Related research includes the work by Li et al.\cite{li2025think}, who explored the application of reinforcement learning in multi-modal large language models (MLLMs) for generative classification. Addressing the phenomenon that models perform better without explicit reasoning in generative classification tasks, they further investigated Adaptive-Thinking and achieved promising results. However, current research still lacks deep investigation into the underlying mechanisms behind the performance improvements brought by reinforcement learning.
Our work systematically explores the performance gains of SFT, DPO\cite{rafailov2023direct}, and GRPO\cite{shao2024deepseekmath} on generative classification tasks under different training and data configurations, while analyzing the underlying causes of these improvements. This provides an effective reference for applying reinforcement learning in generative classification tasks.

\begin{figure*}[!htb]
  \centering 
  \includegraphics[width=0.95\textwidth]{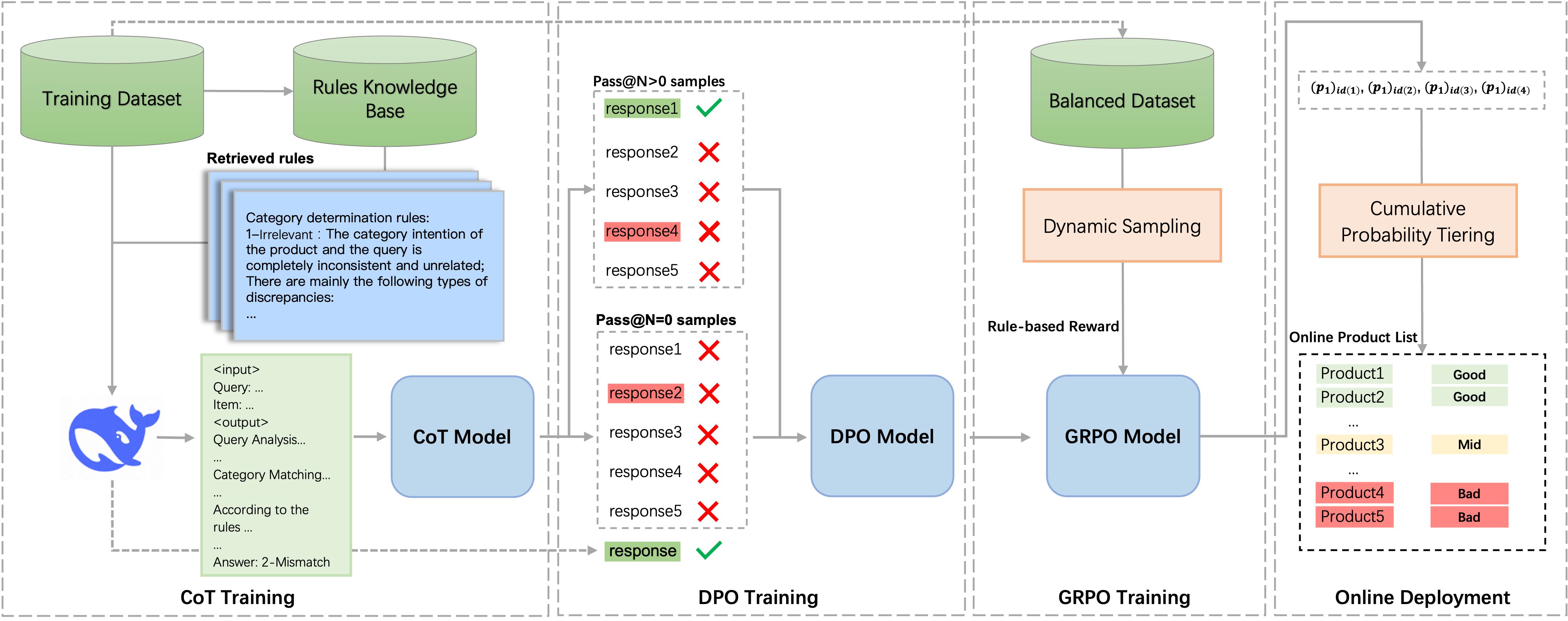}
  \caption{Our TaoSR1 framework comprising three stages: (1) SFT with CoT to endow models with reasoning capabilities; (2) Offline multiple sampling based on a pass@N strategy, combined with DPO, to enhance model generation quality; and (3) Difficulty-based dynamic sampling integrated with GRPO to further mitigate model's discriminative hallucination problems.}
  \Description{Our proposed TaoSR1 framework comprising three stages: (1) SFT with CoT to endow models with reasoning capabilities; (2)Offline multiple sampling based on a pass@N strategy, combined with DPO, to enhance model generation quality; and (3) Difficulty-based dynamic sampling integrated with GRPO to further mitigate model's discriminative hallucination problems.}
\end{figure*}

\section{Methods}\label{sec:methods}
The task of e-commerce search relevance is formulated as a multiclass classification problem, in which the input comprises features of the query and the candidate product, and the output is a predicted relevance label. In the Taobao e-commerce search setting, we adopt a four-tier relevance class scheme: 4-Excellent, 3-Related, 2-Mismatch and 1-Irrelevant.

In this section, we detail our multi-stage optimization framework, which comprises four stages: SFT with CoT, Pass@N-based DPO, GRPO with Difficulty-based Sampling, and Online Deployment. We first perform SFT to endow the model with basic relevance-judgment  capability and introduce a method to extract continuous relevance scores for downstream tiering. To embed nuanced, domain-specific decision logic and prevent the model from degenerating into a purely discriminative model, we integrate CoT during SFT. CoT samples are synthesized via a Retrieval-Augmented Generation (RAG) pipeline to effectively incorporate complex business rules into training. Subsequently, motivated by the model's pass@N accuracy metric, we construct large-scale chosen/rejected pairs for DPO from two disjoint subsets: instances with pass@N accuracy > 0 and instances with pass@N accuracy = 0. The goal is twofold: steer the model to correctly solve instances it is potentially capable of solving, and inject knowledge from a stronger model for the hard cases it consistently fails. 
Finally, informed by analysis of data distribution, we incorporate GRPO with online difficulty-based sampling, which yields further improvements even with small-batch data. Case studies indicate that small-scale online reinforcement learning mitigates discriminative hallucination and improves the quality of the model’s reasoning process.

\subsection{SFT}
\subsubsection{SFT}
Conventionally, training data for query-product relevance prediction consists of <query, item, label> triplets. A discriminative model is fine-tuned on a pre-trained BERT backbone in a point-wise manner, formulated as either a classification or regression task. The corresponding objective functions are typically formulated as follows:
For regression to a single score $\hat{y}$:
\begin{align} 
    \hat{y} =& \mathbf{W}_{\text{mse}}^T \pi(\mathbf{x})  \label{eq:y_hat}\\
    \mathcal{L}_{\mathrm{MSE}}=&\mathbb{E}_{(\mathbf{x}, y) \sim \mathcal{D}}\left[(y-\hat{y})^2\right]
    \label{eq:mse}
\end{align}
For classification to a probability vector $\mathbf{p}$
\begin{align} 
    \mathbf{p} &=\text{Softmax} \left( \mathbf{W}_{\text{ce}} \pi(\mathbf{x}) \right)
    \label{eq:p_i} \\
    \mathcal{L}_{\text{CE}} &=\mathbb{E}_{(\mathbf{x}, \mathbf{y}) \sim \mathcal{D}}\left[-\sum_{c=1}^C y_c \log \left(\mathbf{p})\right)\right]
    \label{eq:ce}
\end{align}
where $C$ is the total number of categories, $\pi(\mathbf{x})$ denotes the model's output representation (e.g., the [CLS] token embedding or a pooled output of all tokens), and $\mathbf{{W}}_{\text{mse}}$ or $\mathbf{{W}}_{\text{ce}}$ represents a learnable projection matrix. For online serving, the resulting scalar $\hat{y}$ or probability vector $\mathbf{p}$ is used in downstream applications. More recently, this paradigm has been extended to Large Language Models (LLMs), which are now frequently adopted as the base model for SFT. In this context, $\pi(\mathbf{x})$ is typically derived from the hidden state of the final End-of-Sequence (EOS) token.
Following large-scale data SFT, both BERT-based and LLM-based approaches yield effective relevance discrimination models, serving as strong baselines. However, training an LLM solely on such discriminative samples effectively collapses an LLM into a pure discriminator, undermining its inherent generative capabilities and misaligning with its original pretraining objective. This prevents the model from fully leveraging the knowledge acquired during pretraining. Consequently, we adopt a generative training paradigm in which the model directly generates the label text (e.g., ``4-Excellent'') (Eq.\ref{eq:gene_ce}). For downstream applications, we extract the probability of the first generated token as a continuous-valued score (Eq.\ref{eq:gene_p}). Formally, this is expressed as:

\begin{equation}\label{eq:gene_ce}
\mathcal{L}_{lm}(\pi)=\mathbb{E}_{(\mathbf{x},\mathbf{y}) \sim \mathcal{D}}\left[-\sum_{t=1}^{|\mathbf{y}|-1} \log \pi\left(\mathbf{y}_{t+1} \mid \mathbf{x}_1, \ldots, y_t\right)\right]
\end{equation}

\begin{equation}\label{eq:gene_p}
\mathbf{P}\left(\mathbf{y}_1=\textbf{c} \mid \mathbf{x}\right)=\left(\mathbf{p}_1\right)_{\text{id(\textbf{c})}}=\frac{\exp \left(\pi(\mathbf{x}\right)_{\text{id(\textbf{c})}})}{\sum_{j=1}^{4} \exp \left(\pi(\mathbf{x}\right)_{{\text{id}(\textbf{c}_j)}})} 
\end{equation}
where $\left(\mathbf{p}_1\right)_{\text{id(\textbf{c})}} \in \mathbb{R}^4$, $\text{id(\textbf{c})}$ denotes the vocabulary index for the special tokens, here $\textbf{c}$ includes [``1'', ``2'', ``3'', ``4'']; $\mathbf{x}_t$ denotes the $t$-th token in a text sequence $\mathbf{x}$; $\mathbf{y}_1$ denotes the extracted first generated token, and its probability is represented by $\mathbf{p}_1$; $\left(\mathbf{p}_1\right)_{\text{id(\textbf{c})}}$ can be converted into a scalar by Eq.\ref{transform} or directly used by downstream task. We will discuss it in detail in Sec.~\ref{ssec:cpt}.

\subsubsection{RAG-based CoT Generation}
Nevertheless, e-commerce relevance annotations are governed by intricate, domain-specific business rules. A single label is insufficient to convey this underlying logic to the model. Simply switching to a generative objective still only leads the model to fit the empirical label distribution of the training data. To address this, we explicitly incorporate a CoT process during SFT to instill domain-specific reasoning capabilities into the model. Our structured CoT framework consists of five sequential steps: (1) Query Understanding, (2) Product Comprehension, (3) Category Matching, (4) Attribute Matching, and (5) Relevance Determination. Using the powerful Deepseek-R1 \cite{deepseekai2025deepseekr1incentivizingreasoningcapability}, we synthesized CoT samples based on our labeled data. Moreover, business rules in e-commerce relevance are complicated and customized. For example, retrieving “Mate50 Pro” for the query “Mate50” is considered 3-Related, whereas retrieving “Mate50” for the query “Mate50 Pro” is rated 2-Mismatch. We therefore aim to enable LLM to more faithfully internalize the business rules in our scenarios. 

During CoT sample synthesis, we incorporate business annotation rules into the prompt and require the model to explicitly cite the applied rules in the final step. However, the full documentation of these rules is extensive. Incorporating it entirely would create impractically long contexts and increase the model's comprehension burden. To circumvent this limitation, we establish a structured knowledge base by decomposing the complete rule set into a collection of fine-grained, ``atomic'' rules, which are indexed by relevant metadata (e.g., category, brand, style) to facilitate efficient retrieval.  During human annotation, we additionally label the atomic factors pertinent to each sample. When constructing CoT samples, we use the atomic factors for a given instance to dynamically retrieve the corresponding rules from the knowledge base and incorporate them into the prompt. This entire procedure constitutes a Retrieval-Augmented Generation (RAG) pipeline. More formally, for each training sample <query, item, label, $\text{reason}_i$>, we use its atomic reasons to dynamically retrieve the corresponding rules from the RAG repository. These retrieved rules are then injected into the prompt, and Deepseek-R1 is prompted with <prompt, RAG-rule, query, item, label> to synthesize the CoT training data. This pipeline yields the final training data in the form <query, item, CoT, label>.

\subsubsection{SFT with CoT}
We experimented with two training paradigms: (1) a ``think-then-respond'' approach, where the model generates the CoT before the label (<CoT, label>), and (2) a ``respond-then-think'' approach with an output format of <label, CoT>. Counterintuitively, the ``think-then-respond'' strategy, despite its theoretical advantages, consistently failed to outperform the discriminative baseline. Our analysis suggests this is due to error accumulation: while CoT generation elevates the model's performance ceiling, showing encouraging results on challenging long-tail cases (e.g., ``alternatives to Miu Miu'', where discriminative models often mishandle it by retrieving Miu Miu itself, failing to capture the semantics of ``alternatives''), the longer generation process is more susceptible to error accumulation. Minor hallucinations or reasoning deviations in the intermediate steps accumulate and ultimately corrupt the final prediction. In contrast, the ``respond-then-think'' setting mitigates this issue by predicting the label first, leading to significantly improved metrics. Nevertheless, a performance gap relative to the discriminative baseline still remained. This leads us to hypothesize that CoT-based SFT, while effective at instilling rule-based logic, cannot generalize well across all scenarios. We posit that subsequent reinforcement learning is essential to unlock model potential.

\subsection{Pass@N-based DPO}

\begin{table}[!tb]
\centering
\caption{Pass@N sampling Results}
\label{tab:passn_sampling}
\setlength{\tabcolsep}{15pt}
\begin{tabular}{lr}
\toprule
Models & Accuracy \\
\midrule
LLM base & 75.01 \\
TaoSR1(CoT) Pass@1 & 67.38\\
TaoSR1(CoT) Pass@2 & 74.26 \\
TaoSR1(CoT) Pass@3 & 77.68\\
TaoSR1(CoT) Pass@4 & 80.18 \\
TaoSR1(CoT) Pass@5 & 81.73 \\
\bottomrule
\end{tabular}
\end{table}
Our offline experimental analysis further corroborates the aforementioned hypothesis. Specifically, we find that the model's pass@N accuracy, derived from multi-pass offline sampling, significantly exceeds that of single-pass decoding. Furthermore, as N increases, this metric eventually surpasses the performance of the pure discriminative baseline. Table \ref{tab:passn_sampling} provides details of these pass@N results. These results suggest substantial room for improvement through reinforcement learning.
We begin by optimizing the model using DPO, leveraging the results of our offline pass@N sampling on training set to construct a preference dataset. This process comprises two stages for constructing the preference dataset:

\begin{itemize}
\item {\texttt{Self-Correction for Solvable Cases}}: For instances where the model produces at least one correct response (i.e., pass@N > 0), we construct preference pairs. A correct response from the sampled candidates is designated as the ``chosen'' ($y^+$), and a randomly selected incorrect response is designated as the ``rejected'' ($y^-$). This process yields a preference dataset $\mathcal{D}_{\text{pass}}$, derived from a subset of our original training set. 
\item {\texttt{Oracle-Guided Correction for Hard Cases}}: A key limitation of self-correction is its inability to address cases the model consistently fails to solve. These hard cases with pass@N $=0$, representing model's consistent failures, are the most critical targets for optimization. Therefore, for this cohort, we sample from DeepSeek-R1 and designate its correct outputs as ``chosen'' responses. These oracle-generated responses are then paired with our model's incorrect ``rejected'' responses to form supplementary preference dataset $\mathcal{D}'_{pass}$. Through this method, approximately $50$\% of the original pass@N $=0$ instances were curated into this new set. 

\end{itemize}
The model is subsequently trained by minimizing the DPO loss over the combined preference datasets $\mathcal{D_{\text{pass}}} \cup \mathcal{D_{\text{pass}}}'$. 
\begin{align}
   \mathcal{L}_{\mathrm{DPO}}\left(\pi_\theta ; \pi_{\mathrm{ref}}\right)&=-\mathbb{E}_{\left(\mathbf{x}, \mathbf{y}_w, \mathbf{y}_l\right) \sim \mathcal{D_{\text{pass}}} \cup \mathcal{D_{\text{pass}}}'} \\
   &\left[\log \sigma\left(\beta \log \frac{\pi_\theta\left(\mathbf{y}^+ \mid \mathbf{x}\right)}{\pi_{\mathrm{ref}}\left(\mathbf{y}^+ \mid \mathbf{x}\right)}-\beta \log \frac{\pi_\theta\left(\mathbf{y}^- \mid \mathbf{x}\right)}{\pi_{\mathrm{ref}}\left(\mathbf{y}^- \mid \mathbf{x}\right)}\right)\right] 
\end{align}

\subsection{GRPO with Difficulty-based Sampling}
The self-correction component of our DPO strategy is conceptually analogous to GRPO, as both encourage models to explore correct reasoning trajectories through multiple sampling of the model's own outputs. For the chosen responses in our pass@N$=0$ set, the label-conditioned synthesis process often induced discriminative hallucination, where the generated rationale contradicts the final results. Relative to DPO, GRPO allows for online sampling with larger N, facilitates more diverse exploration to discover more logically coherent reasoning paths. This makes it a natural subsequent step for further optimization.
Inspired by DAPO \cite{yu2025dapoopensourcellmreinforcement}, we propose a difficulty-aware dynamic sampling strategy. A key departure from DAPO is our handling of homogeneous batches: when a sampled set of responses is all correct or all incorrect, we discard the sample entirely rather than resampling. This is because for query-item pairs with definitive relevance relationships, if multiple sampling still fails to produce correct answers, it indicates that the model may lack requisite knowledge. In such cases, the probability of obtaining a correct answer through further sampling becomes extremely low, which would instead slow down overall training efficiency. For groups in which all responses are correct, the within-group relative advantage is zero, leading to vanishing gradients that also degrade training efficiency. Moreover, during online RL we aim to focus on difficult instances; therefore, we compute advantages and backpropagate gradients only for groups whose empirical accuracy lies within the range $(0,\gamma)$. The training objective is defined as:
\small
\begin{align}
\mathcal{L}_{\mathrm{GRPO}}(\theta)& =\mathbb{E}_{(x,y) \sim \mathcal{D},\left\{o_i\right\}_{i=1}^G \sim \mathcal{D}} \nonumber \\ 
& \left[\frac{1}{G} \sum_{i=1}^G \frac{1}{\left|o_i\right|} \sum_{t=1}^{o_i} \min \left(r_{i, t}(\theta) \hat{A}_t, \operatorname{clip}\left(r_{i, t}(\theta), 1-\epsilon, 1+\epsilon\right) \hat{A}_t\right)\right. \nonumber \\
&\left.-\beta D_{\mathrm{KL}}\left(\pi_\theta \| \pi_{r e f}\right)\right] \nonumber \\
\text { s.t. } 0 & <\mid\left\{o_i \mid \operatorname { is_equivalent }\left(y, o_i\right)\right\} \mid<\gamma \
\end{align}
where 
\begin{equation}
    r_{i, t}(\theta)=\frac{\pi_\theta\left(o_{i, t} \mid x, o_{i,<t}\right)}{\pi_{\theta_{o l d}}\left(o_{i, t} \mid x, o_{i,<t}\right)}, \quad \hat{A}_{i, t}=\frac{R_i-\operatorname{mean}\left(\left\{R_i\right\}_{i=1}^G\right)}{\operatorname{std}\left(\left\{R_i\right\}_{i=1}^G\right)}
\end{equation}
, $R$ is the verifiable reward, indicating whether the final predicted relevance class is correct. The interval configuration for difficulty-based sampling directly determine which data the model emphasizes. Our experiments with various $\gamma$ settings, as reported in Table \ref{tab:difficulty_performance}, showed substantial variation in performance. Further analysis revealed a crucial insight: we identified a strong inverse correlation between the model's final performance and the coefficient of variation (CV) of the label distribution in the GRPO training data. This indicates that a more balanced label distribution yields superior performance on the relevance classification task. This insight motivated a data-centric optimization strategy. Specifically, we curated a balanced subset from the training data by downsampling all majority classes to match the size of the smallest. We then applied a minimal-rejection sampling policy, discarding only homogeneous batches (i.e., $\gamma$ lies in $[0.01,0.99]$). The final experimental results demonstrate that this approach, despite using significantly less data, outperforms GRPO trained on the full, imbalanced dataset. Furthermore, a qualitative analysis of sampled outputs reveals that the incidence of errors attributable to ``discriminative hallucination'' in its failure cases is 30\% lower compared to that of the DPO model.

\subsection{Cumulative Probability Tiering}\label{ssec:cpt}
In the Taobao production search system, the model's relevance predictions are operationalized through a three-tier stratification system: Good, Mid, and Bad, which correspond to our four-point annotation scale (4-Excellent and 3-Related correspond to ``Good'', 2-Mismatch corresponds to ``Mid'', 1-Irrelevant corresponds to ``Bad''). Products are ultimately ranked in descending order of Good, Mid, and Bad tiers.
In online applications, we usually do not directly use the raw classification output, but instead employ a threshold-based mechanism to assign items to these tiers. This allows us to precisely control the precision-recall trade-off based on specific business objectives. For instance, a high-precision setting prioritizes showing only highly relevant items, while a high-recall setting aims to capture a wider set of potentially relevant products. 
Conventionally, models trained with MSE loss typically introduce two hyperparameters as score anchor points for each class: usually 1-Irrelevant is fixed to 0, 4-Excellent is fixed to 1, and the remaining two class's anchor $\alpha_1,\alpha_2$ requires manual adjustment. Two subsequent thresholds $\beta_1,\beta_2$ are then applied to produce the three tiers. For models trained with cross-entropy loss, four hyperparameters are similarly introduced. By computing a weighted average of the probabilities of the first generated class tokens ($\left(\mathbf{p}_1\right)_{\text{id}(\textbf{c})}$, where $\textbf{c} $= [``1'', ``2'', ``3'', ``4'']), we obtain a continuous score via Eq.\ref{transform}, followed by two thresholds $\beta_1,\beta_2$ are used to achieve the three-tier division (See formula \ref{level}).

\begin{equation} \label{transform}
    score = 0*\left(\mathbf{p}_1\right)_{\text{id(1)}} + \alpha_1*\left(\mathbf{p}_1\right)_{\text{id(2)}} + \alpha_2*\left(\mathbf{p}_1\right)_{\text{id(3)}} + 1*\left(\mathbf{p}_1\right)_{\text{id(4)}}  \\
\end{equation}

\begin{equation} \label{level}
\text{rel\_tier} =
\begin{cases}
  \text{good} & \text{if } \text{score}\ge\beta_2 \\
  \text{mid} &  \text{if }  \beta_1 \leq \text{score} < \beta_2 \\
  \text{bad} &
  \text{otherwise}
\end{cases}
\end{equation}
Both conventional methods, therefore, share a significant drawback: they necessitate multiple hyperparameters (at least four), which introduces considerable operational complexity in both training and deployment. Furthermore, the manual tuning of these hyperparameters via grid search is computationally expensive and often yields suboptimal configurations. 
To overcome these limitations, we propose \textbf{Cum}ulative \textbf{P}robability \textbf{T}iering (CumPT), a novel method that extends the standard practice from binary classification---using the positive class probability as the scoring threshold---to the multi-tier setting. CumPT elegantly unifies the process, requiring only a single hyperparameter for online multi-tier assignment, thereby eliminating the complex calibration and thresholding steps.
Under the CumPT framework, model training remains unchanged, utilizing the standard cross-entropy objective (Eq.\ref{eq:gene_ce}) to predict probabilities for each of the four relevance classes. 
This mechanism operates by sequentially accumulating probabilities in descending order of the class tokens (from 4 down to 1) and comparing the sum against a single threshold $\beta_{\text{cum}}$. For instance, if the probability of 4-Excellent $\left(\mathbf{p}_1\right)_{\text{id(4)}}$ or cumulative probability $\left(\mathbf{p}_1\right)_{\text{id(4)}}+\left(\mathbf{p}_1\right)_{\text{id(3)}}$ exceeds $\beta_{\text{cum}}$, the item is assigned to the ``Good'' tier. If not, we evaluate the cumulative probability $\left(\mathbf{p}_1\right)_{\text{id(4)}}+\left(\mathbf{p}_1\right)_{\text{id(3)}}+\left(\mathbf{p}_1\right)_{\text{id(2)}}$; if this sum exceeds $\beta_{\text{cum}}$, the item is assigned to the ``Mid'' tier. If this condition also fails, the item is assigned to ``Bad'' tier. This procedure is formally outlined in Algorithm \ref{alg:cumulative_grading}.

\begin{algorithm}[!ht]
    \caption{Cumulative Probability Tiering} 
    \label{alg:cumulative_grading}          
    \begin{algorithmic}[1] 

        \Require Cumulative probabilities $\left(\mathbf{p}_1\right)_\text{id(\textbf{c})}$ and threshold $  \beta_{cum}$
        \Ensure Relevance tier $\text{rel\_tier}$;

        \If{$\left(\mathbf{p}_1\right)_{\text{id(4)}} \ge \beta_{cum}$}
            \State $\text{rel\_tier} \gets \text{Good}$
        \ElsIf{$\left(\mathbf{p}_1\right)_{\text{id(4)}}+\left(\mathbf{p}_1\right)_{\text{id(3)}} \ge \beta_{cum}$}
            \State $\text{rel\_tier} \gets \text{Good}$
        \ElsIf{$\left(\mathbf{p}_1\right)_{\text{id(4)}}+\left(\mathbf{p}_1\right)_{\text{id(3)}}+\left(\mathbf{p}_1\right)_{\text{id(2)}} \ge \beta_{cum}$}
            \State $\text{rel\_tier} \gets \text{Mid}$
        \Else
            \State $\text{rel\_tier} \gets \text{Bad}$
        \EndIf
        
        \State \textbf{return} $\text{rel\_tier}$
        
    \end{algorithmic}
\end{algorithm}

\begin{table*}[!tb]
\setlength{\tabcolsep}{15pt}
\caption{Offline Evaluation Results}
\label{tab:all_performance}
\centering
\resizebox{\textwidth}{!}{
\begin{tabular}{lrrrrrr}
\toprule
Models & Class-1 F1 & Class-2 F1 & Class-3 F1 & Class-4 F1 & Macro F1 & Accuracy \\
\midrule
BERT & 65.74 & 69.63 & 33.87 & 76.06 &61.33 & 69.36\\
Qwen3-0.6B & 42.02 & 68.13 & 23.50 & 78.14 & 52.95 & 70.29\\
Qwen3-30B-A3B &65.09 & 68.80 & 32.47& 81.68 & 62.01 & 74.42 \\
LLM-base & 65.19 & 68.86 & 32.91 & 81.90 & 62.22 & 75.04\\
\midrule
TaoSR1(CoT)  & 43.30 & 67.54 & 19.68 & 75.62 & 51.54 & 68.22\\
TaoSR1(CoT\&DPO) & 62.90 & 71.20 & 37.96 & 82.25 & 63.58 & 75.54\\
TaoSR1(CoT\&DPO\&GRPO) & 66.82 & 72.15 & 39.40 & 82.27 & 65.16 & 75.92\\
\midrule
TaoSR1(CoT)$_{\text{postCoT}}$ & 57.63 & 72.64 & 27.91 & 81.88 & 60.01 & 75.12\\
TaoSR1(CoT\&DPO)$_{\text{postCoT}}$ &65.74 & 71.95 & 39.43& 83.00 & 65.03 & 76.49\\
TaoSR1(CoT\&DPO\&GRPO)$_{\text{postCoT}}$ & \textbf{67.34} & \textbf{73.15} & \textbf{44.94} & \textbf{83.06} & \textbf{67.12} & \textbf{76.86} \\
\bottomrule
\end{tabular}
}
\end{table*}

\begin{table}[!tb]
\centering
\caption{Dataset Statistics}
\label{tab:label_distribution}
\setlength{\tabcolsep}{20pt} 
\begin{tabular}{lrr} 
\toprule
Label & Count & Ratio \\
\midrule
L4 &38,739 & 50\%\\
L3 & 3,863 & 5\% \\
L2 &27,913 & 36\% \\
L1 & 6,865 & 9\% \\
\bottomrule
\end{tabular}
\end{table}

\begin{table}[!tb]
\centering
\caption{Pass@5 Statistics}
\label{tab:pass_rate}
\resizebox{0.9\linewidth}{!}{
\begin{tabular}{lcccccc}
\toprule
Pass@5 &0 & 1 &2 & 3 &4 & 5 \\
\midrule
Ratio & 13.7\% & 4.9\% & 4.6\% &5.3\% & 7.4\% & 64.1\% \\
\bottomrule
\end{tabular}
}
\end{table}

\begin{table}[!tb]
\centering
\caption{Impact of Difficulty Ratio}
\label{tab:difficulty_performance}
\resizebox{\columnwidth}{!}{
\begin{tabular}{lcccccc}
\toprule
Difficulty Ratio & L1& L2 & L3& L4 & L-cv & Macro F1 \\
\midrule
{[}0.2-0.8{]} & 3.6\% & 29.1\% & 6.4\% &60.9\% & 1.06 & 63.55 \\
{[}0.2-0.6{]} & 4.6\% &29.9\% & 8.2\% & 57.3\% & 0.97 & 63.87\\
{[}0.1-0.5{]} & 6.0\% & 31.8\% &10.5\% & 51.7\% & 0.84 & 64.25\\
\midrule
\makecell[l]{{[}0.01-0.9{]} \\ w balance label} & 25.0\% & 25.0\% &25.0\% & 25.0\% & 0& 67.12 \\
\bottomrule
\end{tabular}
}
\end{table}

\begin{table}[!tb]
\centering
\caption{Ablation of DPO}
\label{tab:dpo_ablation}
\setlength{\tabcolsep}{15pt}
\begin{tabular}{lc}
\toprule
Models & Macro F1\\
\midrule
TaoSR1(CoT)$_\text{postCoT}$ & 60.01\\
TaoSR1(CoT\&GRPO)$_\text{postCoT}$ & 66.84 \\
TaoSR1(CoT\&DPO)$_\text{postCoT}$ & 65.02 \\
TaoSR1(CoT\&DPO\&GRPO)$_\text{postCoT}$ & \textbf{67.12} \\
\bottomrule
\end{tabular}
\end{table}

\section{Experiments}
\subsection{Experiment Setup}
\subsubsection{Dataset and Metrics}
Our test set is collected from Taobao's online search logs, sampled from the scoring candidate space of the relevance model. It comprises about 70,000 manually annotated query-item pairs. To rigorously evaluate the model's reasoning capabilities, the query distribution is intentionally focused on four challenging categories: negation, affordable alternatives, question-answering (QA), and knowledge-based queries.
The overall test set label distribution is shown in Table~\ref{tab:label_distribution}. It can be observed that for the relevance model's scoring candidates, after filtering through the preceding pipeline with relevance factors, the proportion of completely irrelevant (1-Irrelevant) samples has become quite small. However, the more challenging partially irrelevant (2-Mismatch) samples still account for a relatively large proportion. Overall, the dataset maintains a relatively balanced distribution between relevant (i.e., 4-Excellent and 3-Related combined) and non-relevant (2-Mismatch and 1-Irrelevant combined) instances.

For offline metrics, we evaluate our model using macro-F1 and compare F1 metrics for each tier. For online evaluation, we conduct live A/B experiments and rely on human assessment to compute the following key metrics:

\begin{itemize}
\item {\texttt{GSB (Good/Same/Bad)}}: This is a side-by-side comparative framework where human assessors compare results returned by two methods for the same query in A/B experiments. GSB measuring the relative superiority of the test bucket compared to the baseline bucket. GSB$+x\%$ can be understood as $x\%$ of results in the test bucket being better than the base bucket.
\item {\texttt{Query Goodrate}}: This is a page-level relevance metric. The entire search results page for a query is classified into ``Good'', ``Mid'', or ``Bad'' tiers based on the proportion of relevant items displayed. The metric itself is the proportion of queries whose pages are rated as ``Good'' or ``Mid''.
\item {\texttt{Item Goodrate}}: This is an item-level metric that measures the absolute proportion of highly relevant items (i.e., 4-Excellent, 3-Related) per request. The item goodrate for this evaluation is the average of item goodrates across all requests.
\end{itemize}

It is important to note the distinction between these metrics: Query Goodrate and Item Goodrate measure absolute performance levels, for which we report the absolute lift. In contrast, GSB captures the relative advantage of our proposed model, for which we report the relative improvement.

\subsubsection{Baselines}
We employ multiple models as our baselines, including BERT, Qwen3-0.6B, and Qwen3-30BA3B\cite{yang2025qwen3technicalreport}. Our model is iteratively developed based on Tbstar. Tbstar is a closed-source e-commerce large language model foundation trained from scratch by Taobao, utilizing over $10$T tokens of general corpus and rich domain-specific e-commerce corpus. Based on a Mixture of Experts (MoE) architecture, it has a total parameter count of $42$B with $3.5$B activated parameters.
\begin{itemize}
\item {\texttt{BERT}}: A 24-layer BERT model foundation trained from scratch based on general corpus and Taobao's internal e-commerce corpus, subsequently fine-tuned on the full training set using formula \ref{eq:ce}.
\item {\texttt{Qwen3-0.6B}}: The 0.6B parameter open-source dense model from the Qwen family. It was supervised fine-tuned (SFT) on our complete training dataset using the formula \ref{eq:gene_ce}.
\item {\texttt{Qwen3-30BA3B}}: An open-source Mixture-of-Experts (MoE) model from the Qwen family, featuring 30 billion total parameters and 3 billion active parameters per forward pass. This model was also supervised fine-tunedon our full training set same as Qwen3-0.6B.
\item {\texttt{LLM base}}: This serves as our strongest baseline and represents the starting point for our further optimizations. It is the Tbstar-42B model after being supervised fine-tuned (SFT) on our full training dataset using the formula \ref{eq:gene_ce}.
\item {\texttt{TaoSR1}}: This represents our full, multi-stage optimization pipeline built upon the LLM Base model, incorporating the CoT, DPO, and GRPO phases as described in Sec.~\ref{sec:methods}. We specifically report results for TaoSR1$_{\text{postCoT}}$, which denotes our final model employing the ``respond-then-think'' architecture for efficient deployment.
\end{itemize}

\begin{table*}[!t]
\centering
\caption{Performance Comparison of Different Methods with Threshold Tuning}
\label{tab:method_threshold_comparison}
\begin{tabular}{llrrrrrr}
\toprule
Methods & threshold & Macro F1 & Online Macro F1 & Good F1 & Good Precision & Good Recall \\
\midrule
\multirow{5}{*}{Formula-based} & $\beta_2$ = 0.3 &67.12 & 41.42 & 73.66& 58.73 & 98.77 \\              & $\beta_2$ = 0.4 & 67.12 & 64.25 & 85.29 & 79.81 & 91.59\\
                              & $\beta_2$ = 0.5 & 67.12 & 64.83& 85.34 &80.43 & 90.89 \\
                              & $\beta_2$ = 0.6 & 67.12 & 65.26 & 85.36 & 80.87 & 90.37 \\
                              & $\beta_2$ = 0.7 & 67.12& 56.83 &83.35 & 81.05 & 85.78 \\
\midrule
\multirow{5}{*}{CumPT} & $\beta_{\text{cum}}$ = 0.3 & 67.12 & 67.05 & 85.37 & 80.50 & 90.88 \\
                       & $\beta_{\text{cum}}$ = 0.4 & 67.12& 67.13 &85.39 & 80.75 & 90.59\\
                       & $\beta_{\text{cum}}$ = 0.5 & 67.12 & 67.14 & 85.42 & 81.04 & 90.30 \\
                       & $\beta_{\text{cum}}$ = 0.6 & 67.12 & 67.15& 85.40 &81.26 & 89.99 \\
                       & $\beta_{\text{cum}}$ = 0.7 & 67.12 & 67.17 & 85.39 & 81.51 & 89.66\\
\bottomrule
\end{tabular}
\end{table*}

\begin{table*}[!htb]
\centering
\caption{Side-by-Side Human Evaluations}
\label{tab:sbs}
\setlength{\tabcolsep}{8pt}
\begin{tabular}{p{2cm}p{6cm}ccc}
\toprule
Query Type & Case & GSB & Query Goodrate & Item Goodrate \\
\midrule
Q\&A & What medicine can make hair black? & +16.62\% & +6.53pt & +5.66pt \\
Alternative & Miumiu alternative & +34.43\% & +13.11pt & +10.69pt \\
Negative & Short sleeves that don't stick to hair & +10.92\% & +3.80pt & +3.74pt \\
Knowledge & Paint that is not afraid of car pressure & +18.45\% & +6.85pt & +4.44pt \\
\bottomrule
\end{tabular}
\end{table*}

\subsection{Offline Evaluation}
Table \ref{tab:all_performance} presents a comprehensive comparison between our model and various baselines. We focus on the macro-F1 metric rather than accuracy, as it provides a balanced measure of classification performance that is insensitive to class imbalance. Our results reveal three key observations:

\textbf{Domain-specific pre-training is crucial}. The24-layer BERT pre-trained on our in-house e-commerce corpus significantly outperforms the Qwen3-0.6B model. Similarly, LLM base shows a substantial performance gain over the comparably sized Qwen-30BA3B.

\textbf{Directly incorporating CoT in SFT degrades classification performance due to error propagation}. This is evident in the TaoSR1(CoT) model, which shows a marked drop in F1-scores across all classes compared to the LLM Base. In contrast, our TaoSR1(CoT)$_{\text{postCoT}}$ model, which adopts the ``respond-then-think'' architecture, circumvents this issue by predicting the label before generating the CoT. This leads to a significant performance recovery over the standard CoT approach.

\textbf{Reinforcement learning unlocks full potential of CoT for downstream classification task}. TaoSR1(CoT\&DPO) achieves substantial performance improvements compared to TaoSR1(CoT) and also surpasses LLM base. Similarly, TaoSR1(CoT\&DPO)$_{\text{postCoT}}$ also demonstrates superiority compared to TaoSR1(CoT)$_{\text{postCoT}}$ and LLM base. Further optimization with GRPO yields additional performance gains, indicating that larger sampling spaces are key for reinforcement learning to enhance model capabilities.

Ultimately, our model achieves optimal performance under the post-CoT configuration through reinforcement learning optimization, with a $4.9$ pt point improvement in macro-F1 compared to LLM base. TaoSR1$_{\text{postCoT}}$ not only resolves the error accumulation problem but also makes online deployment of the model feasible.

\subsection{Ablation Study}
\subsubsection{Ablation of Post-CoT}
We contrast the ``think-then-respond'' paradigm with the ``respond-then-think'' paradigm (distinguished by the postCoT subscript) in Table \ref{tab:all_performance}. We evaluate the performance of each paradigm both at the SFT stage and subsequent DPO stage. At the SFT-only stage, the post-CoT approach demonstrates a significant advantage over the standard approach. This provides strong evidence that post-CoT effectively mitigates the issue of error accumulation in CoT, positioning it as a potentially superior approach for classification tasks. Furthermore, the application of DPO consistently yields notable performance gains under both configurations. This confirms that even within the post-CoT framework, there remains substantial headroom for improvement through reinforcement learning.

\subsubsection{Why DPO before GRPO}
A natural question arises regarding the necessity of a distinct DPO stage, given that its core mechanism for pass@N > 0 samples---constructing preference pairs from offline multi-pass sampling---is conceptually analogous to the online sampling in GRPO.
The key reason for the DPO stage lies in its handling of ``hard cases'' where our model consistently fails (i.e., pass@N = 0). For this samples, self-sampling is insufficient. Instead of relying on our model's flawed outputs, we leverage a more powerful, general-purpose ``oracle'' model to generate the ``chosen'' responses. These are then paired with our model's own incorrect ``rejected'' responses.
Through this approach, DPO training not only optimizes the model's own reasoning paths but also introduces incremental information from external models. This beneficial information cannot be obtained solely through sampling the model itself.
The value of this intermediate DPO stage is empirically validated by our experiments (see Table \ref{tab:dpo_ablation}). These results demonstrate that initiating the GRPO phase from a DPO-refined checkpoint yields superior offline performance compared to applying GRPO directly to the SFT-only baseline.

\subsubsection{Ablation of CumPT}
This section provides a detailed analysis of advantages of our CumPT method. Macro F1 is computed directly on model's raw outputs, representing its intrinsic classification capability. Online Macro F1 is computed after the tiering mentioned in Sec.~\ref{ssec:cpt} (either the traditional formula-based approach via Eq.\ref{transform}, Eq.\ref{level} or CumPT). Additionally, we report Precision, Recall, and F1-score specifically for ``Good'' tier.

To illustrate the sensitivity of the traditional method, we fixed a subset of its many hyperparameters ($\alpha_1=0.33$, $\alpha_2=0.66$, $\beta_1=0.15$) and varied only a single parameter $\beta_2$ for comparison. The results in Table \ref{tab:method_threshold_comparison} reveal two critical findings: (1) CumPT achieves robust and stable performance with a single hyperparameter. Notably, the resulting Online Macro F1 score closely tracks the original Macro F1, indicating minimal performance degradation during the tiering process. (2) The traditional method exhibits high sensitivity to its hyperparameter settings, leading to significant variance in performance. This highlights a practical challenge that identifying the optimal combination of its multiple hyperparameters requires extensive and costly experimentation.

In summary, CumPT not only simplifies the deployment process by reducing the number of required hyperparameters but also delivers more consistent and predictable performance, effectively bridging the gap between a model's offline potential and its online efficacy.

\subsection{Online Evaluation}
To validate the effectiveness of the model in real online applications, we first conducted a side-by-side evaluation based on manual annotation. We selected 2,000 queries, requested both experimental buckets for each query, and compared the top-10 results across three metrics. The specific results are shown in Table \ref{tab:sbs}.
We provide a fine-grained analysis of these results by segmenting the queries into four distinct categories. The most substantial gains were observed for queries seeking affordable alternatives. This is because such queries pose a classic challenge for traditional term-matching models, which often fail to distinguish between a target brand and its alternatives. For instance, a query for a ``Miumiu alternative'' would often cause a term-based model to incorrectly retrieve authentic Miumiu products with high relevance scores due to strong lexical overlap. Furthermore, significant improvements were also recorded across the other query categories. These results collectively demonstrate that the reasoning and general knowledge capabilities of our LLM-based model can effectively address long-standing challenges in search relevance --particularly for queries that require a deep, semantic understanding of user intent rather than simple lexical matching.

In addition, we conducted strict online A/B tests to evaluate changes in user purchase-related behaviors. The primary metrics include Item Page View (IPV), Transaction volume, and Gross Merchandise Value (GMV). The results show a  0.22\% increase in UV, 2.43\% increase in IPV, a 0.82\% increase in Transaction volume, and a 0.29\% decrease in GMV. Overall, these metrics remain essentially on par with those of the baseline bucket, indicating that the optimization of our relevance model does not impair users’ purchase intent while substantially improving their overall shopping experience.




\section{Conclusion}
In this paper, we introduce a comprehensive framework for optimizing large language models for search relevance. Our methodology begins by endowing the model with structured reasoning capabilities through SFT with CoT.  We leverage a Retrieval-Augmented Generation (RAG) pipeline to synthesize the CoT samples, effectively integrating complex business logic into the training process. Crucially, we explore both a conventional ``think-then-respond''  structure and a novel ``respond-then-think'' paradigm. The latter allows the model to internalize sophisticated reasoning patterns while remaining compatible with the low-latency demands of real-time online deployment. Furthermore, we employ a multi-stage reinforcement learning pipeline, incorporating DPO and GRPO, to significantly enhance the model's robustness on challenging instances. This complete framework demonstrates significant and consistent improvements in both offline and online evaluations. The proposed framework offers a new paradigm for developing and deploying LLM-based relevance models. Moreover, our findings provide valuable insights into the application of LLMs for classification tasks, particularly on how to balance advanced reasoning capabilities with practical system constraints.

\bibliographystyle{ACM-Reference-Format}
\bibliography{sample-base}

\appendix

\section{Appendix}

\subsection{Implementation Details}
During the Supervised Fine-Tuning (SFT) phase, we used a batch size of $1024$ and a learning rate of $1e-6$, managed by a cosine scheduler with a warmup ratio of $0.05$. We employed the AdamW \cite{loshchilov2018decoupled} optimizer with $\beta_1=0.9$ and $\beta_2= 0.999$. The model was trained for $2$ epochs with a maximum input length of $4096$ tokens. 

For Direct Preference Optimization (DPO), we maintained a batch size of $1024$ and a learning rate of $1e-6$, again using a cosine scheduler with a $0.05$ warmup ratio and the AdamW optimizer. The training proceeded for $500$ steps with the maximum input length kept at $4096$ tokens. To stabilize training, we incorporated an auxiliary SFT loss term applied specifically to the ``chosen'' responses, with a weighting coefficient of $0.5$. The pass@N samples are generated with hyperparameters top-k as $20$, top-p as $0.95$, and temperature as $0.6$. A detailed statistics of the pass@N data distribution is provided in Table \ref{tab:pass_rate}.

In the GRPO reinforcement learning phase, we configured a rollout batch size of $64$ with a learning rate of $1e-6$. For each sample, we generated $16$ responses using a sampling temperature of $0.99$ and top-k of $100$. The thresholds for our online difficulty sampling were set to a range of $[0.01, 0.9]$. For stabilization, we clipped the importance sampling ratio to a range of $1\pm0.2$ and the advantage values to $\pm2$. Our training pipeline is implemented based on the open-source framework ROLL\cite{wang2025reinforcementlearningoptimizationlargescale}.

\subsection{Online Deployment}
For online deployment, we directly served the TaoSR1 model in the production environment. Due to its relatively large size, the initial request Response Time (RT) was high, prompting us to implement several engineering optimizations to improve online throughput.
\begin{itemize}
\item{\texttt{Optimizing load balance.}}
Instead of using a fixed batch size, we partition incoming requests according to their sequence length, ensuring that the computational workload across batches is approximately balanced.

\item{\texttt{Scheduling across data-centers.}}
We introduced proxy services for data centers in different regions, which poll and dispatch micro-batched requests to idle machines, thereby alleviating cross–data-center scheduling bottlenecks.

\item{\texttt{Refining model deployment.}}
We enabled FP8 quantization for our relevance model, and refactored the key-value cache reuse mechanism to support prefix sharing within the same batch for requests with identical prefixes.
\end{itemize}
After these optimizations, the average online RT per user request decreased from 800 to 286 ms, and the Model FLOPs Utilization (MFU) finally reaches 46.2\%. Taking advantage of the post-CoT generation paradigm adopted in our relevance model, the online system only needs to optimize latency in the prefill phase, without introducing additional overhead from the decoding phase.

\subsection{Further Analysis of Post-CoT}

To further analyze the effectiveness of Post-CoT and prevent it from overfitting to the offline evaluation set—which often leads to "hacking" the metrics—we proposed a more objective evaluation method. Overfitting is a common issue; in our past experiments, we observed that simple tricks like swapping the query and item inputs or repeating the query could increase offline scores but failed to produce any real benefits in online deployment.

We aimed to build a benchmark that is independent of query and relevance grade distributions to objectively evaluate the Post-CoT model. Therefore, we established an \textbf{Atomic Capability Evaluation Benchmarks (v1)} for relevance modeling. This preliminary set allows us to derive initial insights. We divided the relevance judgment process into four dimensions: \textbf{E-commerce General Knowledge}, \textbf{Business Rule Adherence}, \textbf{Reasoning}, and \textbf{Feature Utilization}. Based on our analysis, most online relevance issues fall into one of these four categories.

We defined clear criteria for these capabilities and had annotators label a balanced dataset of 2,000 samples. We used Accuracy (Acc) to evaluate the following models:

\begin{itemize}
    \item \textbf{TaoSR1(CoT)$_{\text{postCoT}}$}: Supervised Fine-Tuning (SFT) using only the labeled data.
    \item \textbf{Qwen3-32B(CoT)$_{\text{postCoT}}$}: SFT on a larger base model using the same labeled data.
    \item \textbf{TaoSR1(CoT\&DPO\&GRPO)$_{\text{postCoT}}$}
    \item \textbf{TaoSR1(CoT\&DPO\&GRPO)$_{\text{postCoT}}$-RAG}: An experimental offline version that adds external RAG information during training and inference.
\end{itemize}

\begin{figure}[ht]
  \centering
  \includegraphics[width=0.85\columnwidth]{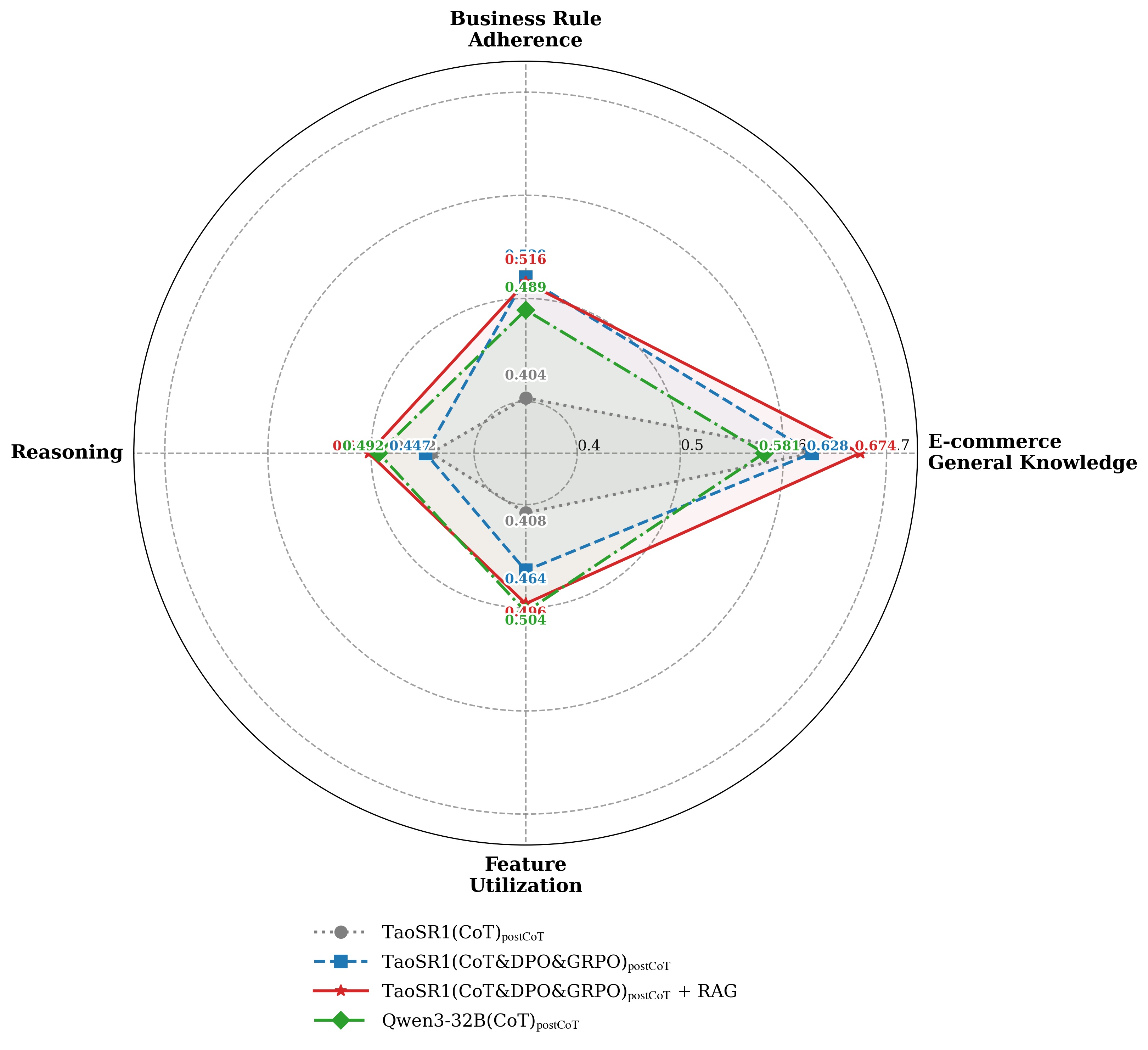}
  \caption{Radar chart of model performance across \textbf{Atomic Capability Evaluation Benchmarks}.}
  \label{fig:radar_chart}
  \Description{Radar chart of model performance across \textbf{Atomic Capability Evaluation Benchmarks}}
\end{figure}

The results are shown in Figure~\ref{fig:radar_chart}. We draw several important conclusions:

\begin{enumerate}
    \item \textbf{Model Scale:} More active parameters lead to overall improvements across all dimensions (shown by the \textbf{Green Line}). However, this creates greater challenges for online performance (e.g., latency).
    
    \item \textbf{E-commerce General Knowledge:} Introducing RAG brings the most significant improvement in the dimension of E-commerce General Knowledge (shown by the \textbf{Red Line}).This demonstrates that RAG is a critical approach for supplementing the model's missing knowledge.
    
    \item \textbf{Reasoning:} Post-CoT does not essentially improve reasoning capabilities. Its gains come primarily from better adherence to business rules and more detailed utilization of features (comparing the \textbf{Blue Line} vs. the \textbf{Grey Line}). 
\end{enumerate}
Based on the current findings, it appears that significant improvements in reasoning capability may still rely heavily on the Pre-CoT paradigm. Consequently, implicit CoT or AdaptiveCoT may represent promising avenues for future exploration. Therefore, finding an optimal way to enhance online reasoning using CoT—while balancing model effectiveness, latency, and resources—remains a significant and challenging avenue for future research.

\end{document}